# Spontaneous formation and optical manipulation of extended polariton condensates

E. Wertz[1], L. Ferrier[1], D. Solnyshkov[2], R. Johne[2], D. Sanvitto[3], A. Lemaître[1], I. Sagnes[1], R. Grousson[4], A.V. Kavokin[5], P. Senellart[1], G. Malpuech[2] and J. Bloch[1*]

[1]*Laboratoire de Photonique et de Nanostructures, LPN/CNRS, Route de Nozay, 91460, Marcoussis, France*

[2]*LASMEA, Clermont Université, Université Blaise Pascal, CNRS, 24 av des Landais, Aubiere, 63177, France*

[3] *Departamento de Fisica de Materiales, Universidad Autonoma de Madrid, Madrid 28049, Spain*

[4] *Institut des NanoSciences de Paris, CNRS UMR 7588, Université P. et M. Curie, Campus Boucicaut, 140 Rue de Lourmel, 75015 Paris, France*

[5] *Physics and Astronomy School, University of Southampton, Highfield, Southampton, SO171BJ, UK*

*\* e-mail : Jacqueline.bloch@lpn.cnrs.fr*

**Cavity exciton-polaritons[1,2] (polaritons) are bosonic quasi-particles offering a unique solid-state system to investigate interacting condensates[3,4,5,6,7,8,9,10]. Up to now, disorder induced localization and short lifetimes [4,6,11] have prevented the establishment of long range off diagonal order[12] needed for any quantum manipulation of the condensate wavefunction. In this work, using a wire microcavity with polariton lifetimes ten times longer than in all previously existing samples, we show that polariton condensates can propagate over macroscopic distances outside the excitation area, while preserving their spontaneous spatial coherence. An extended condensate wave-function builds-up with a degree of spatial coherence larger than 50% over distances 50 times the polariton De Broglie wavelength. The expansion of the condensate is shown to be governed by the repulsive potential induced by photo-generated excitons within the excitation area. The control of this local potential offers a new and versatile method to manipulate extended polariton condensates. As an illustration, we demonstrate synchronization of extended condensates via controlled tunnel coupling[13,14], and localization of condensates in a trap with optically controlled dimensions.**

Modern semiconductor technology allows the realisation of nanostructures where both electronic and photonic states undergo quantum confinement. In particular in semiconductor microcavities, excitons confined in quantum wells and photons confined in a Fabry Perot resonator can enter the light-matter strong coupling regime. This gives rise to the formation of cavity polaritons, mixed exciton-photon states which obey bosonic statistics [1]. The polariton dispersion presents a sharp energy minimum close to the states with zero in-plane wave-vector (k = 0) with an effective mass $m^*$ three orders of magnitude smaller than that of the bare quantum well exciton. Recently polariton Bose-Einstein condensation[3,4,5,6,7,8,9,10] (BEC) and related effects such as vortices [15,16] or superfluid[17,18,19] behavior have been reported at unprecedented high temperatures. Because of their finite lifetime, cavity polaritons are a model system to investigate dynamical BEC[20,21] with a technological control of the resonator geometry and the polariton lifetime. In previously reported polariton BEC systems, the cavity lifetime and the photonic disorder prevented the build-up of extended condensates needed for the realization of polariton circuits [22,23]. The measured coherence length ranged at best from 10 to 20 μm[4,6,11,24] a few times the polariton thermal De Broglie wavelength.

In the present work, we report on the spontaneous formation of extended polariton BEC with a spatial coherence extending over 50 times the thermal De Broglie wavelength. Spatial control of such extended condensates is demonstrated, opening the way to a new range of physical phenomena. A large Rabi splitting (15 meV) microcavity of high optical quality is used, with a cavity photon life time around 15 ps, 5 to 50 times larger than in previous reports (see Methods). To manipulate the condensate wave functions, we use one dimensional cavities defined by 200 µm long microwires with a width between 2 and 4 µm (fig. 1.a). Such a small transverse wire dimension induces a lateral quantum confinement of polaritons and the formation of 1D sub-bands (see fig 1.b for a 3.5 µm wire).

Polaritons are generated optically using a non resonant excitation provided by a single mode continuous wave laser focused down to a 2 µm size spot. The laser energy is detuned from the lower polariton branch towards higher energy (typically by 100 meV). At low excitation power, the far field emission (resolved in k-space) shows polariton population along two 1D branches (fig 1.b). When raising the excitation power, a strong and abrupt increase of the overall wire emission intensity is observed, as shown in fig. 1c. Spatially resolved measurements above threshold (see fig. 2b) show that a polariton condensate is spontaneously formed within the excitation spot at the lowest energy polariton state, close to $k_y = 0$[10]. This condensate undergoes a local blueshift $E_b$ due to repulsive interaction with the high-density exciton cloud photogenerated in the excitation area. Since there are no photogenerated excitons outside the small excitation area, this blueshift is spatially limited and creates a force tending to expulse polaritons from the excitation area: the polariton condensate undergoes a lateral acceleration and acquires a finite in-plane wave-vector[25,26]. The far field emission measured on each side of the excitation spot indicates the spontaneous generation of condensed polaritons with well-defined wave-vectors given by $k_y = \pm \sqrt{2m^* E_b}/\hbar$ when approximating the polariton dispersion by a parabola [26]. As the excitation power is increased, $E_b$ and consequently $k_y$ increases, perfectly following the polariton dispersion (see fig. 2d). The real space image of the wire (figure 2.e) shows that this finite in-plane wave vector leads to the propagation and expansion of the polariton condensate far on both sides of the excitation area. Energy analysis of the emission demonstrates that this extended condensed phase coexists with the cloud of uncondensed polaritons and excitons at higher energy, which remains in the close vicinity of the excitation spot (see the magnified upper panel of fig 2.e).

To evidence the spatial coherence of this extended condensate, we perform a Young slit interferometry [11,27] experiment (fig 3.a). An image of the photonic wire is formed in the plane of a screen, pierced with two tilted slits. The spatial coherence between two small areas (which dimensions are of the order of 0.5 µm) symmetrically located at a distance $a/2$ on each side of the excitation spot is probed. Figures 3.b and 3.c show interference patterns recorded below and above threshold. Below threshold, the emission is spectrally broad and interference fringes are barely visible. In contrast, above threshold pronounced fringes are observed with high visibility even for $a$ as large as 200 µm. The first order spatial coherence $g^1(a)$ can be deduced from theses interference patterns using: $g^1(a) = U(a) \frac{<I_1(r)> + <I_2(r)>}{2\sqrt{<I_1(r)><I_2(r)>}}$ where $\langle I_i(r) \rangle$ is the average intensity measured with a single slit at position $r$ in projection plane. $U(a) = \frac{<I(r)>_{max} - <I(r)>_{min}}{<I(r)>_{max} + <I(r)>_{min}}$ is the contrast between the maximum and minimum average intensities measured with both slits. Figure 3d shows the measured $g^1(a)$ for several excitation powers. Below threshold, the coherence is 20% close to the excitation spot ($a = 5$ µm), and rapidly goes to zero for distances above $a = 20$ µm. Just above threshold, the coherence starts to increase and extends over a larger distance. Finally, for excitation powers exceeding the threshold by a factor of 3 or more, high spatial coherence is found all along the wire with an extracted coherence length of the order of 0.5 $mm$ — much larger than the wire length. This measurement demonstrates the formation of a coherent polariton wave function over the whole wire, a key feature proving the spontaneous build-up of macroscopic off-diagonal long range order (ODLRO) in the whole system[12].

Note that in a 1D system, BEC spatial coherence is expected to be highly sensitive to thermal phase fluctuations. Indeed, in such a system at thermal equilibrium, the two point density matrix characterizing the spatial coherence presents an exponential decay with a characteristic length given by $r_0 = \frac{2n_0\hbar^2}{m^*k_BT}$ [28], where $k_B$ is the Boltzmann constant, $n_0$ the condensate density, and $T$ the temperature. Therefore, a 1D condensate can not present spatial coherence over a distance larger than $r_0$. In our case, $r_0$ is of the order of 1 $mm$, much larger than the wire length which means that possible phase fluctuations do not prevent the establishment of ODLRO as observed here. Further evidence of the coherence is obtained when approaching the laser beam at a distance D from the wire end. An interference pattern between waves traveling back and forth from the wire end is visible in the emission intensity along the wire (see supplementary I).

In the following, we show that under higher excitation powers, several polariton condensates can coexist and be optically manipulated in a 1D cavity. Indeed, at high excitation powers, polariton-polariton interactions become efficient close to the excitation area, and induce polariton scattering toward lower energy states (see supplementary II). Several extended condensates appear on both sides of the excitation spot which acts as a potential barrier (see figure 4.a.). Young slit interferometry performed within either side of the pumping area evidences that each of these condensates shows ODLRO (see figure 4.b.). Various regimes are evidenced when probing the spatial coherence between condensates located on opposite sides from the excitation spot. As shown in figure 4.c., low energy condensates present no spatial coherence between the left and right side of the spot. Indeed, at these lowest energies, the potential barrier is thick enough to prevent tunnel coupling between the two sides. On the contrary, for the highest energy condensates, a small signal in the central barrier evidences the tunnel coupling between the right and left part of the condensate. This tunnel coupling is responsible for phase locking of the two spatially separated condensates[13,14]. As a result, high spatial coherence all along the wire is evidenced for these states (see fig. 4 c). Thus the repulsive potential created by optical pumping under the excitation spot can be used to generate extended condensates synchronized by a controlled tunnel coupling, which opens the way to the realization of polariton Josephson junctions.

The simultaneous observation of condensation on several discrete states is the signature of the dynamical condensation regime in this 1D system: the build up of a macroscopic population for certain states results from the complex interplay between scattering rates, energy conservation and lifetime of the polariton states. Control of these final states can be obtained by approaching the laser spot to the wire end: an energy trap with well defined quantized states can be formed as illustrated in figure 4.d and 4.e. This trap is delimited on one side by the wire end and on the other one by the adjustable barrier defined by the excitation spot. 0D condensates are formed as evidenced by the wave-functions directly imaged in the intensity distribution within the trap.

When the excitation spot is close to the wire end (fig.4.d and 4.e), the potential barrier on the left side simply reflects the Gaussian shape of the excitation spot. Because of efficient relaxation into the trapped states, the high energy condensate does not expand on this side. On the other side, the potential barrier clearly presents a larger bottom part, looking like a pedestal, induced by the efficient expansion of the condensate. Indeed when the spot lies far from both ends, the pedestal on the potential barrier is observed on both sides as seen in fig.4.a.

As compared to previously reported techniques to spatially confine polariton condensates[5,7,29,30], the size and height of the trap can be continuously controlled by changing both the excitation power and the position of the spot. Controlling both the propagation and the wavefunction of a polariton condensate opens the way to the realization of polariton circuits, a first step toward a high speed all optical information processing[22].

Finally we note that the macroscopic propagation we report here is a unique feature of polariton condensates. Indeed we could also obtain regular photon lasing in the same wires at temperatures above 60 K (see supplementary III): in this photon lasing regime, the coherent emission is limited to the excitation area without any propagation feature. On the opposite, gain induced confinement within the excitation spot is observed.

**Methods**

The photonic wires are fabricated using electron-beam lithography and reactive ion etching from the cavity sample described in ref [10]. Microphotoluminescence is measured at 10 K on a single wire using a single mode continuous Ti:Sapphire laser. Far field (respectively near field) spectroscopy is performed by projecting the Fourier plane (respectively the image plane) of the microscope objective used to collect the emission upon the entrance slit of a monochromator. The signal is then detected by a nitrogen-cooled CCD camera. The sample shows a Rabi splitting of 15 meV. The exciton-photon detuning is defined as $\delta = E_C(k_y=0) - E_X(k_y=0)$ where $E_C(k_y=0)$ and $E_X(k_y=0)$ are the cavity mode and the exciton energy for ky = 0.

**Acknowledgement**

This work was partly supported by the C'Nano Ile de France contract "Sophiie2", by the ANR contract PNANO- 07-005 GEMINI, by the FP7 ITN "Clermont4" (235114) and by the FP7 ITN "Spin-Optronics"(237252).

**Figure captions:**

**Figure 1 : Cavity polaritons confined in a mimcrowire cavity**

Scanning electron micrograph of the array of cavity wires; b) Emission of a single wire as a function of detection angle below threshold ($P = 0.2\ Pth$): such far field measurements give direct information about the polariton distribution in $k_y$ space since the emission angle $\theta$ of a polariton state with energy $E$ is linked to $k_y$ by $E\sin(\vartheta) = \hbar c k_y$. Two 1D polariton sub-bands are homogeneously populated; c) Emission intensity measured on the whole wire as a function of the excitation power. For b) and c) $L_x = 3.5\ \mu m$ and $\delta = -3\ meV$

**Figure 2 : Condensation and spatial spreading of cavity polaritons in microwires**

a), c) : Far field emission above threshold (*P = 1.5 Pth*) of a 30 µm wire section located on either side of the excitation area, b) far field emission for *P = 1.5 Pth* of a 10 µm wire section centered on the excitation spot; d) Measured blueshift of the condensate as a function of the condensate wave-vector; also shown the calculated polariton dispersion and photon-like dispersion. e) Real space distribution intensity along the wire measured above threshold (*P = 2.5 Pth*): condensed polaritons spread over the whole wire while higher energy excitons remain in the excitation area (magnified top panel). $L_x = 3.5 \ \mu m$ and $\delta = -3 \ meV$

**Figure 3: Macroscopic off-diagonal long range order**

a) Schematic of the double slit interferometry set-up for spatial coherence measurements: the interference pattern measured at infinity is projected onto the entrance slit of the spectrometer with a lens of focal length *f*; b) interference patterns for a distance *a = 20 µm* between the two slits measured below threshold (P = 0.5 $P_{th}$); c) interference of the emission of the two wire ends (*a = 200 µm*) measured above threshold (P = 10 $P_{th}$); d) First order spatial coherence deduced from the fringe visibility measured for various excitation powers as a function of *a*. Error bars have been estimated considering the intensity fluctuations on the interference patterns. $L_x = 3.5 \ \mu m$ and $\delta = -2 \ meV$

**Figure 4: Manipulation of the condensate wave-function**

We illustrate here how the condensate wave-functions can be optically manipulated using the repulsive potential within the excitation spot. a) Emission intensity measured along the wire for an excitation power *P = 30 $P_{th}$*; polariton relaxation induces the build-up of several delocalized condensates at lower energy; b) and c) Spatial coherence measurements using two slits b) on the same side or c) on each side of the excitation area : tunnel coupling through the central barrier allows synchronization of each side of the condensates for the highest energy states. d) and e) Trapping of the polariton condensates when approaching the laser spot to the wire end (P = 7.5 $P_{th}$). The wavefunctions of the discrete trapped states are directly imaged. The solid lines represent the barrier in the excitation area: we calculate the Coulomb potential induced by photoinjected excitons using a spatial distribution corresponding to the observed potential. a), b) and c) : $L_x$ = 2.5 μm and $δ$ = -3 meV; d) and e): $L_x$ = 3.5 μm and $δ$ = -2 meV

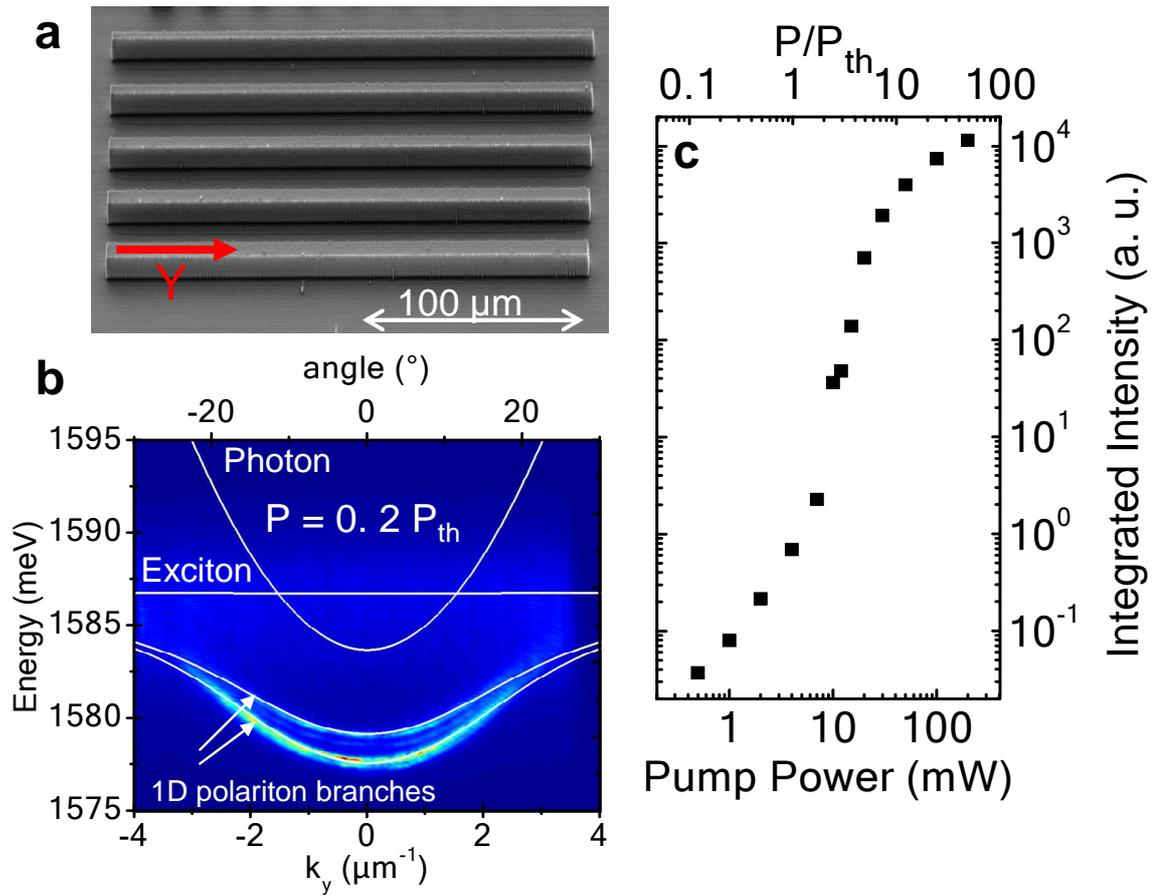

Figure 1

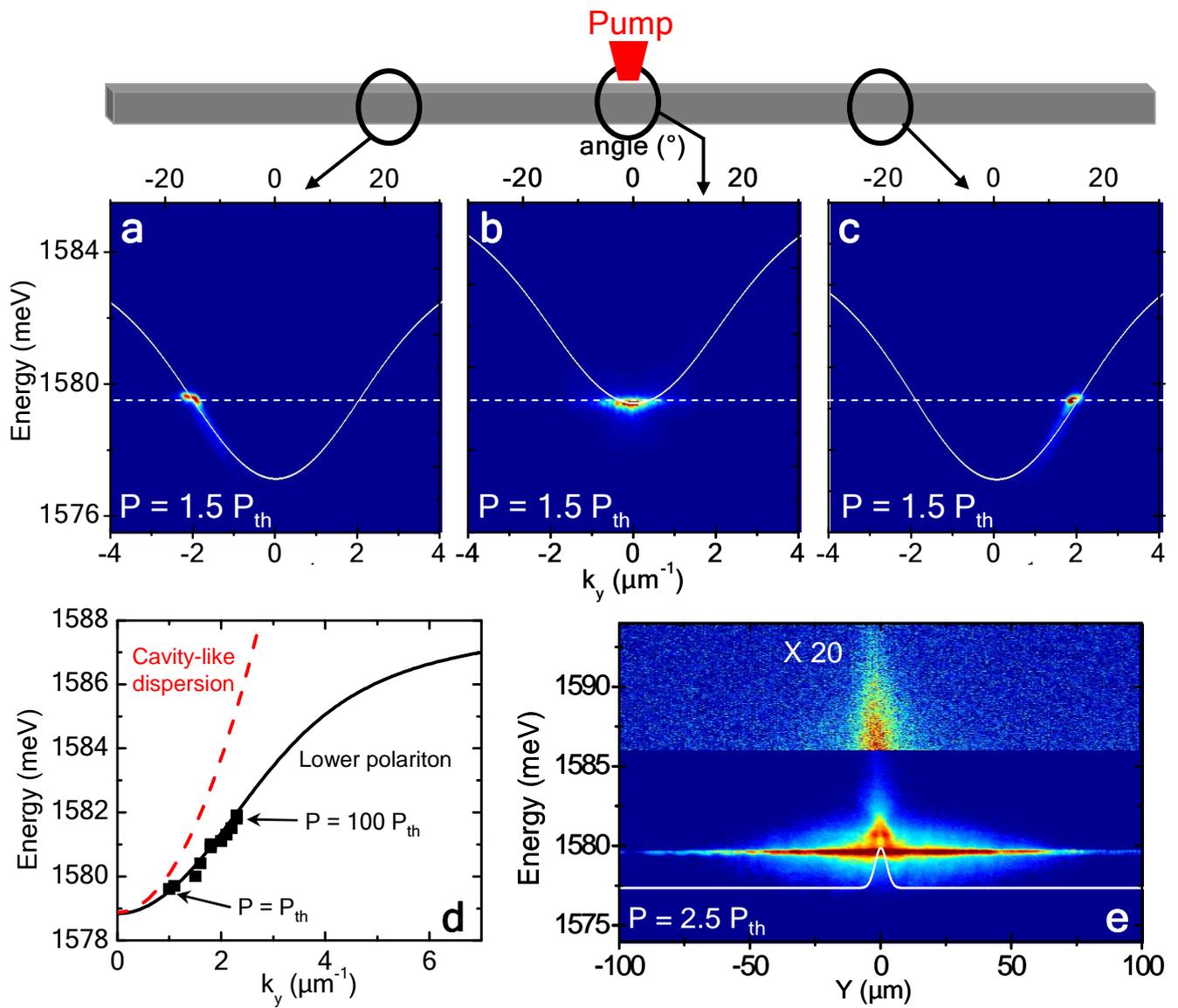

Figure 2

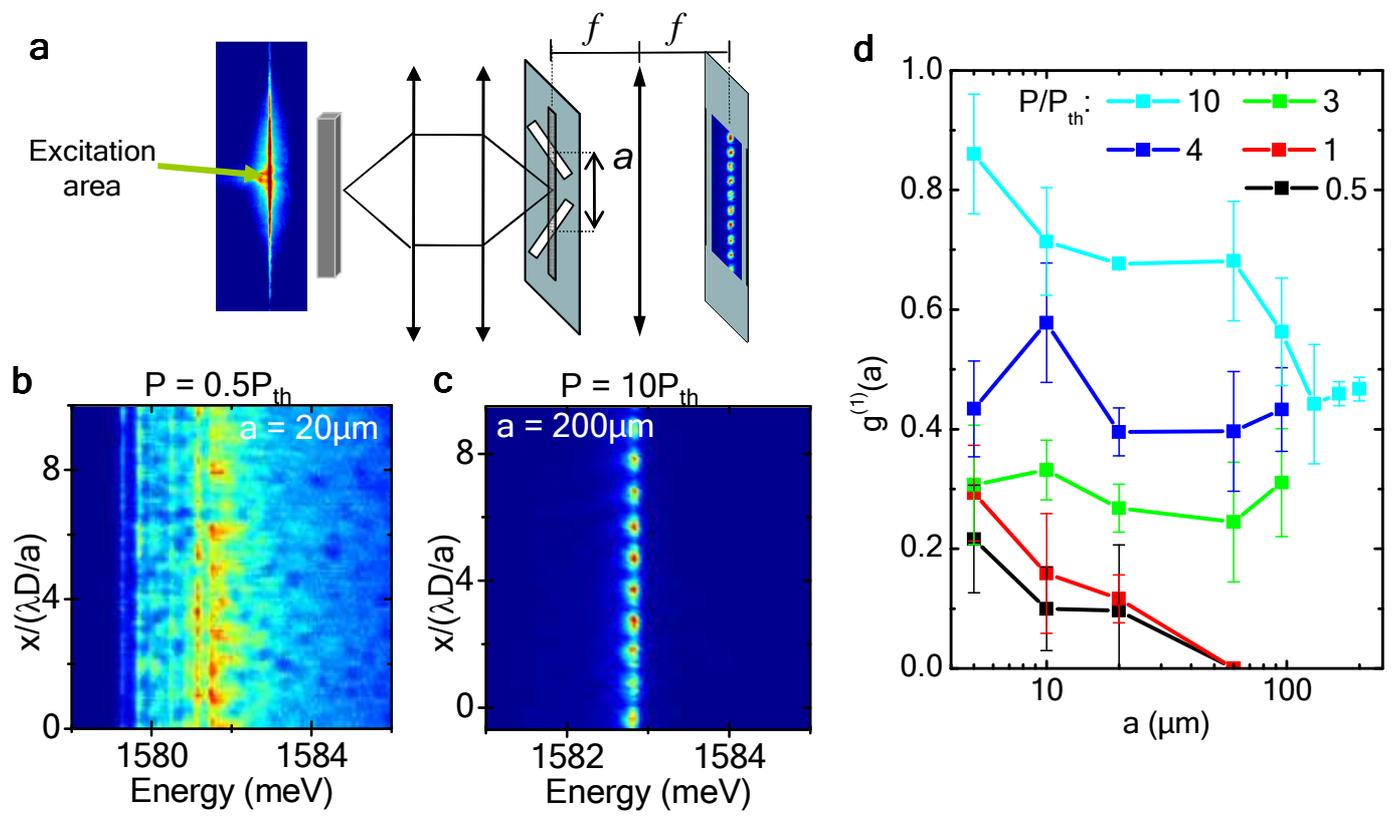

Figure 3

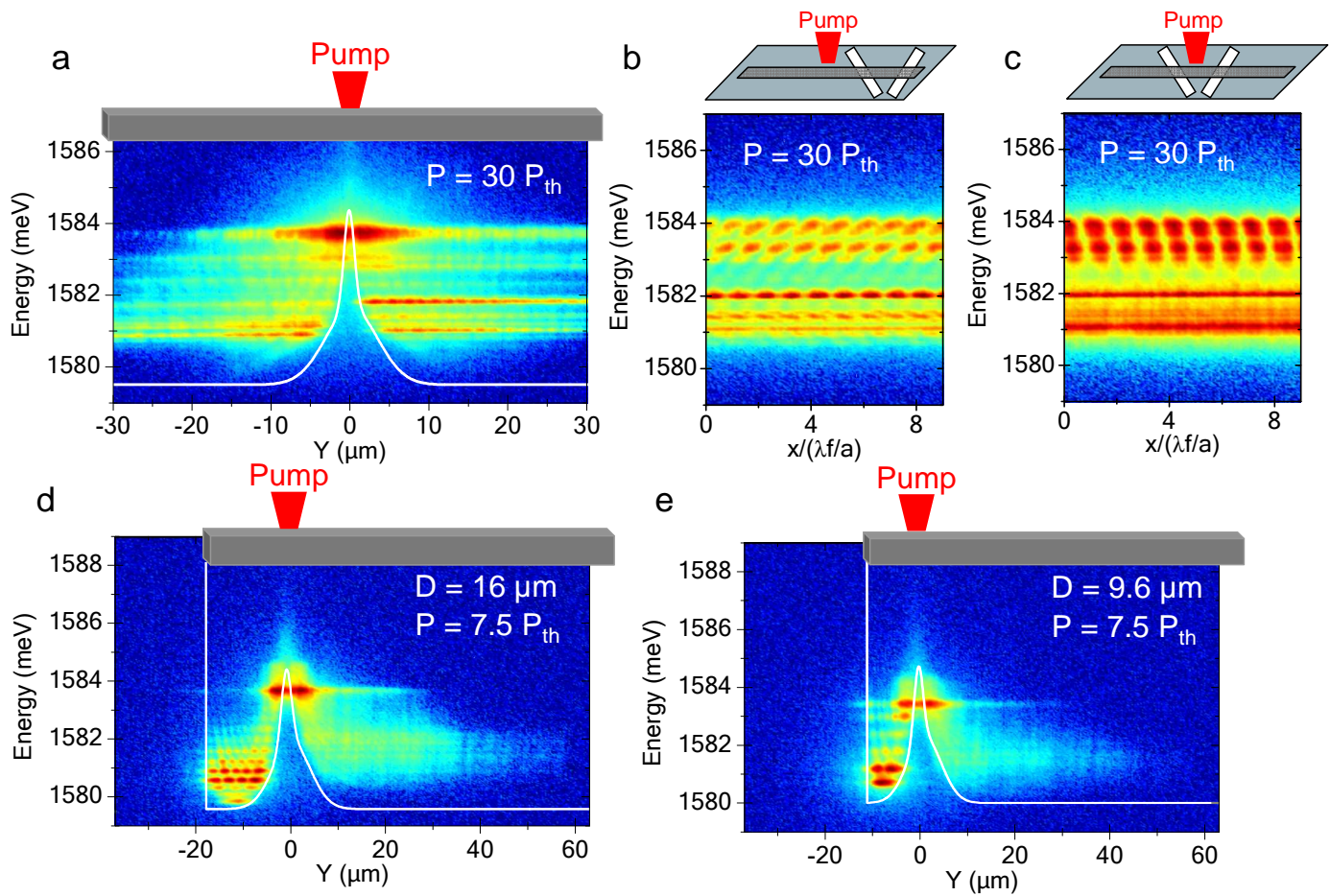

Figure 4

# Supplementary information

I. **Spatial modulation of the condensate wave-function**

Further evidence of the polariton condensate coherence is obtained when approaching the laser beam at a distance D from the wire end. An interference pattern between waves traveling back and forth from the wire end is visible in the emission intensity along the wire (see figures 1a and 1b). Such modulation of the emission intensity along the wire can be theoretically reproduced, solving numerically the time-dependent 1D Schrödinger equation including resonant pumping and finite polariton lifetime. Particles are introduced by a *cw* coherent pumping having a 2 µm size and located at a distance D from the edge of the wire. The repulsive potential created by the localized electron-hole pair distribution is modeled as an external Gaussian potential having the same size as the one of the pump. This model describes a coherent motion of the condensate wave function in the mean-field potential formed by the edge of the wire and by excitons created by non-resonant pumping. A *cw* coherent pumping is included in the model, as well as the pump-induced local repulsive potential. The model neglects dephasing or scattering of the polaritons. The good agreement with the experimental data demonstrates that the decay of the amplitude of the interference fringes versus D is mainly due to the finite particle lifetime and that the dephasing/scattering processes are not efficient on the time scale of the polariton lifetime.

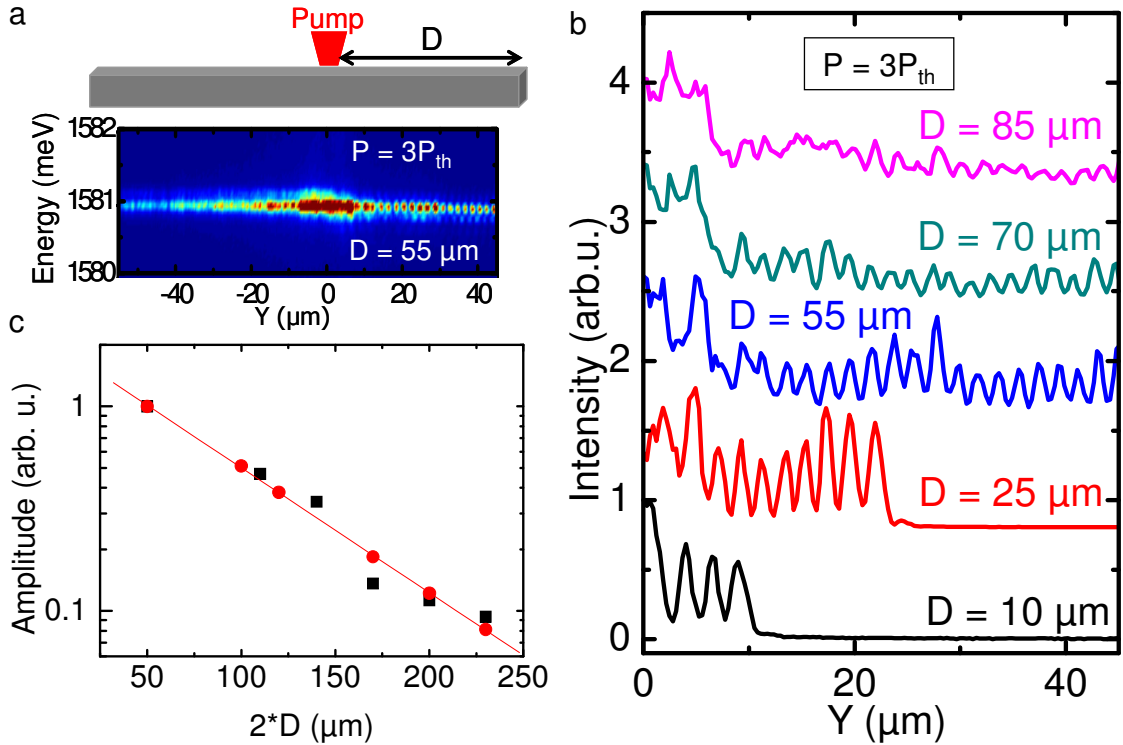

FigureA:

a) Emission intensity measured along the wire for a distance $D = 55\ \mu m$ between the excitation spot and the wire end; b) Intensity profiles measured along the wire at the energy of the polariton condensate for various values of $D$; c) Amplitude of the interference fringes as a function of $D$: the black squares represent the experimental data and the red dots a fit, calculated using the Schrödinger equation including a cw coherent generation of polaritons at the pumping location, a trapping potential, and a 35 ps polariton lifetime. The red line is a linear fit, guide for the eye. $L_x = 3.5\ \mu m$ and $\delta = -4.5\ meV$.

## II. Behaviour under higher excitation conditions: polariton relaxation and photon lasing

We illustrate here how polariton relaxation progressively becomes more efficient as the excitation power is increased. Several extended polariton condensates can be generated on each side of the excitation area. The left column of figure B corresponds to the excitation spot positioned at the centre of a wire with $L_x = 3.5\ \mu m$ and $\delta = -3\ meV$. On the right column, the spot lies 20 µm from the wire end and a trap is formed between the wire end and the excitation spot. In this experiment $L_x = 3.5\ \mu m$ and $\delta = -5\ meV$.

Notice that at the highest excitation powers, the excitation area can enter the weak coupling regime. However outside the excitation spot, the system remains free from high energy carriers and extended polariton condensates are still injected.

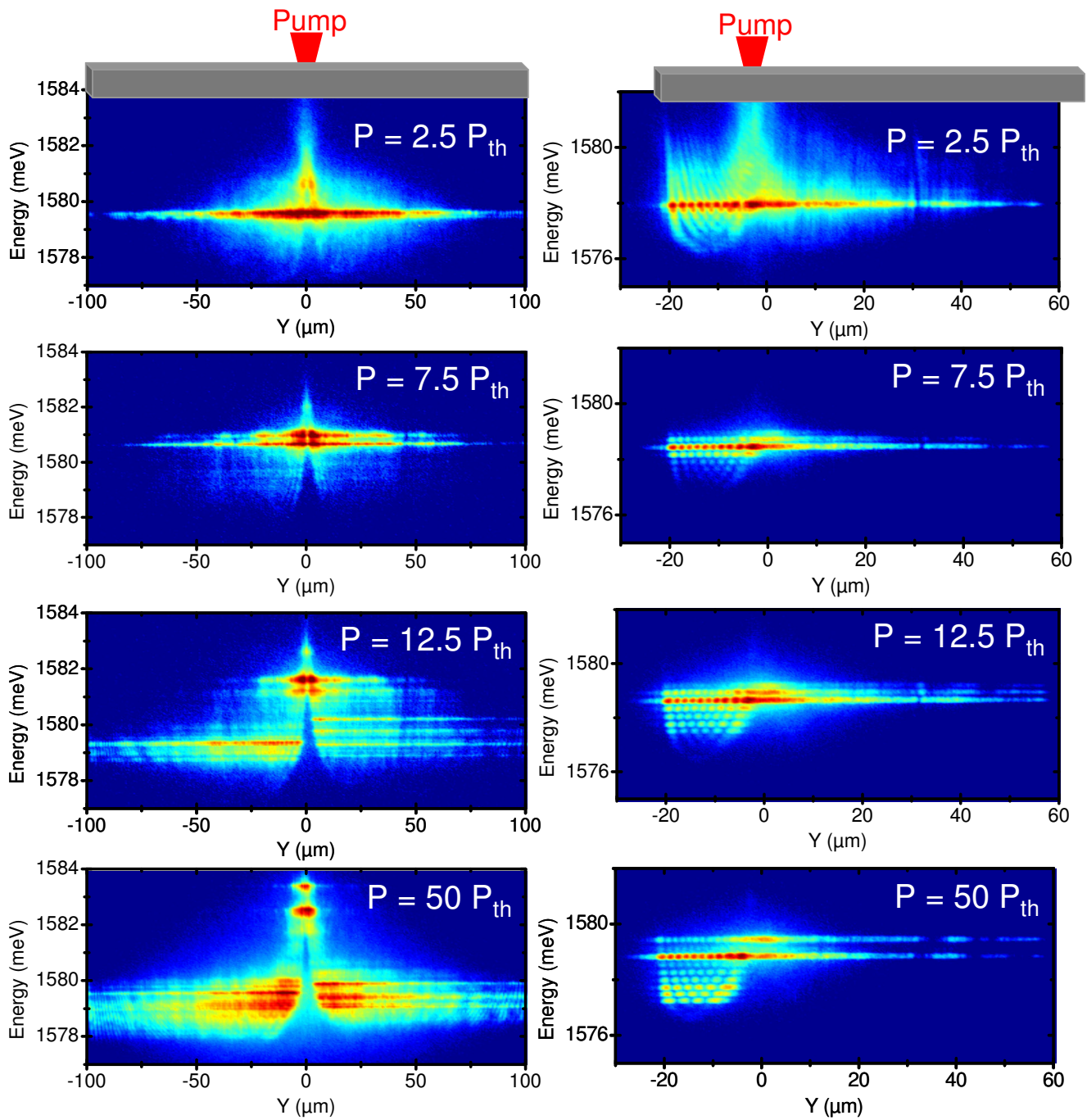

Figure B

## III. Comparison with regular photon lasing : Behaviour at higher temperature

Polariton condensation in GaAs microcavities is observed only at temperatures below 50K, as predicted theoretically [i]. This is illustrated in fig.C.a. At 100 K, the polariton dispersion has indeed vanished around threshold and is replaced by a dispersion characteristic for the cavity mode. Real space imaging above threshold shows that the coherent emission in the photon lasing regime is restricted to the excitation area. No long distance propagation is observed as opposed to the polaritonic regime at lower temperature. $L_x = 3.5~\mu m$ and $\delta = -2~meV$

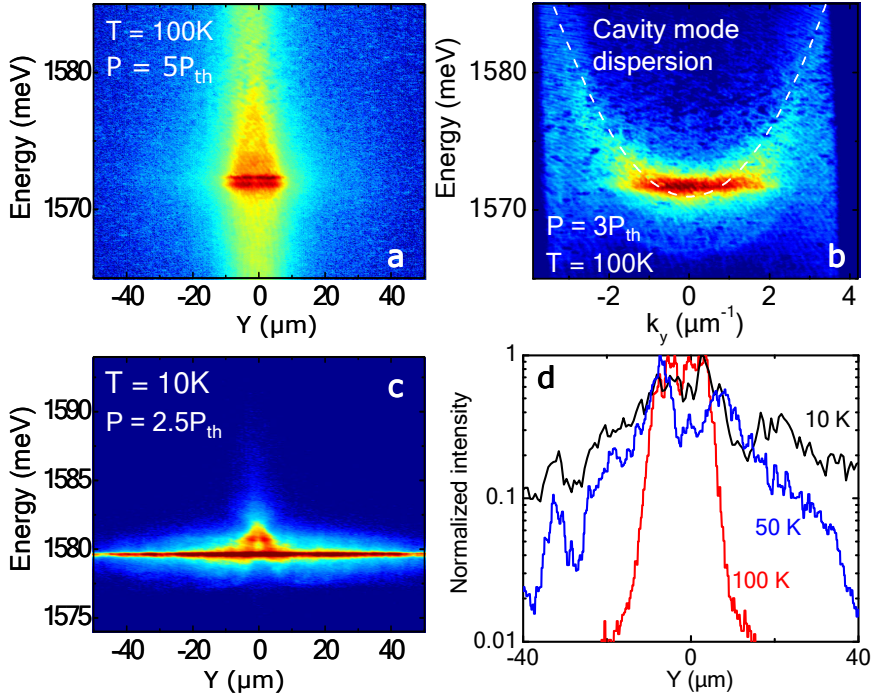

Figure C : a) Emission intensity measured along the wire at 100K above threshold for photon lasing $(P = 5P_{th})$ ; b) Far field emission measured above photon lasing threshold $(P = 3P_{th})$ at 100K: the characteristic cavity mode dispersion is observed; c) Emission intensity measured along the wire above polariton condensation threshold $(P = 2.5P_{th})$ at 10K ; d) Intensity profiles measured along the wire above threshold at different temperatures. $L_x = 3.5~\mu m$ and $\delta = -3~meV$ in all measurements.

Reference :
[i] Malpuech, G. *et al.*, Polariton laser : Thermodynamics and quantum kinetic theory. *Semicond. Sci. Technol.*, **18**, S395, (2003)